\documentclass[aip,graphicx]{revtex4-1}
\usepackage{siunitx}
\usepackage{graphicx}
\usepackage{amsmath}
\usepackage{bm}

\draft 

\begin{document}


\title{Magnon Bose-Einstein condensates: from time crystals and quantum chromodynamics to vortex sensing and cosmology} 



\author{J.T. M\"akinen}
\email[]{jere.makinen@aalto.fi}
\affiliation{Low Temperature Laboratory, Department of Applied Physics, Aalto University, POB 15100, FI-00076 AALTO, Finland}

\author{S. Autti}
\email[]{s.autti@lancaster.ac.uk}
\affiliation{Department of Physics, Lancaster University, Lancaster LA1 4YB, UK}

\author{V.B. Eltsov}
\email[]{vladimir.eltsov@aalto.fi}
\affiliation{Low Temperature Laboratory, Department of Applied Physics, Aalto University, POB 15100, FI-00076 AALTO, Finland}



\date{\today}

\begin{abstract}
Under suitable experimental conditions collective spin-wave excitations, magnons, form a Bose-Einstein condensate (BEC) where the spins precess with a globally coherent phase. Bose-Einstein condensation of magnons has been reported in a few systems, including superfluid phases of $^3$He, solid state systems such as Yttrium-iron-garnet (YIG) films, and cold atomic gases. Among these systems, the superfluid phases of $^3$He provide a nearly ideal test bench for coherent magnon physics owing to experimentally proven spin superfluidity, the long lifetime of the magnon condensate, and the versatility of the accessible phenomena. We first briefly recap the properties of the different magnon BEC systems, with focus on superfluid $^3$He. The main body of this review summarizes recent advances in application of magnon BEC as a laboratory to study basic physical phenomena connecting to diverse areas from particle physics and cosmology to new phases of condensed matter. This line of research complements the ongoing efforts to utilize magnon BECs as probes and components for potentially room-temperature quantum devices. In conclusion, we provide a roadmap for future directions in the field of applications of magnon BEC to fundamental research.
\end{abstract}


\maketitle 


\section{Introduction}
Spin waves are a general feature of magnetic materials. Their quanta are called magnons, spin-1 quasiparticles that obey bosonic statistics. At sufficiently large number density and low temperature, magnons form a Bose-Einstein condensate (BEC), akin to neutral atoms in superfluid $^4$He or ultracold gases. The magnon BEC is manifested as spontaneous coherence of spin precession across a macroscopic ensemble in both frequency and phase even in the presence of incohering forces resulting from, e.g., magnetic field gradients. 

Bose-Einstein condensation of magnons (in the form of a {\it homogeneously precessing domain}, HPD) was first observed in 1984 \cite{HPD1} in nuclear magnetic resonance (NMR) experiments in the superfluid B phase of $^3$He. The experiments discovered a spontaneous formation of coherent precession of magnetization within a superfluid with antiferromagnetic ground state. In the early experiments, spontaneous coherence of magnons was manifested in pulsed NMR measurements\cite{HPDexp} in a very peculiar manner. At first the amplitude of the free induction decay signal dropped, as expected for a dephasing set of local oscillators in an inhomogeneous magnetic field. However, shortly after the signal amplitude was restored and, even more surprisingly, the observed ringing time was a few orders of magnitude longer than the initial dephasing time! These observations were explained using the terminology of spin superfluidity that acted as the mechanism establishing spontaneous coherence. 

Later experiments demonstrated spin supercurrent between two homogeneously precessing domains connected by a channel \cite{BRcurrent}, and showcased the spin current analog of the DC Josephson effect \cite{BR2}. Further signatures of spin superfluidity and macroscopic coherence include the observation of topological objects, spin vortices \cite{BRvortex}, and the collective Goldstone modes, i.e. oscillations of the phase of precession \cite{Goldstone3HeB, PhysRevLett.100.155301}. These modes can be treated as phonons in a time crystal, discussed in Sec.~\ref{sec:PhononsInTC}.

\begin{figure}
\includegraphics[width=0.4\linewidth]{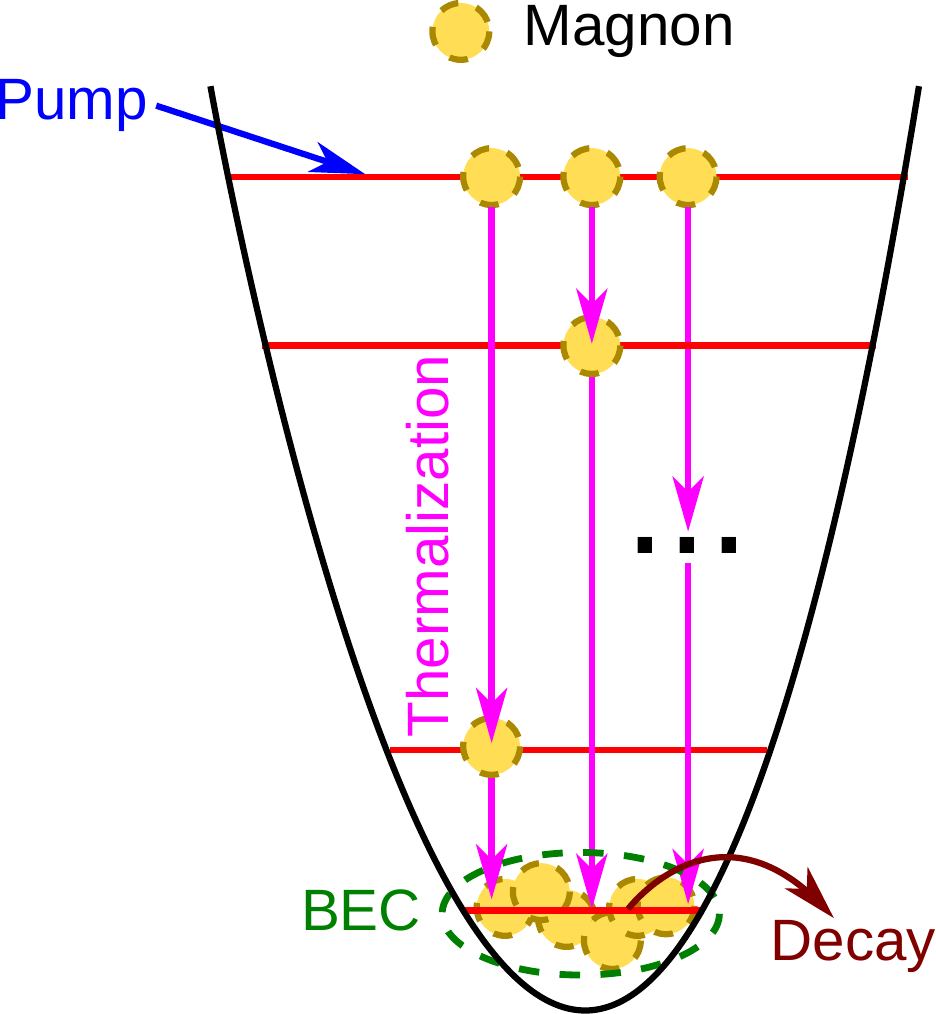}
\caption{{\bf Creation of the magnon BEC.} In a typical scheme quanta of spin-wave excitations (magnons) are pumped into a higher energy level by e.g. a radio-frequency pulse. The pumped magnons then thermalize with time constant $\tau_{\rm E}$. The magnon decay from the ground state is characterized by the decay time $\tau_{\rm N}$. Under sufficiently strong pumping, or if the magnon decay time is much longer than the thermalization time, $\tau_{\rm N} \gg \tau_{\rm E}$, a macroscopic number of magnons occupy the ground state of the system, forming a BEC.}
\label{Fig:MagnonScheme}
\end{figure}

Treating the coherent spin precession as a Bose-Einstein condensate of magnons allows for a simplified description of the system. The condensate wave function amplitude describes the density of magnons and the phase the precession of magnetization. The magnon BEC differs from the atomic BECs in one important respect: magnons are quasparticles and, thus, their number is not conserved. In a thermodynamic equilibrium, magnon chemical potential is always zero, $\mu_{\rm eq} \equiv 0$, and thus no equilibrium BEC of magnons can exist. If the lifetime of magnons $\tau_{\rm N}$ is much larger than the thermalization time $\tau_{\rm E}$ within the magnon subsystem, i.e., $\tau_{\rm N} \gg \tau_{\rm E}$, or if magnons are continuously pumped into the system, the magnon condensate obtains a nonzero $\mu$ and forms a lowest-energy-level condensate analogous to a BEC as illustrated in Fig.~\ref{Fig:MagnonScheme}. These conditions are well met e.g. in $^3$He-B, where the thermalization time is a fraction of a second while the lifetime of the free coherent precession and the corresponding magnon BEC can reach tens of minutes at the lowest temperatures \cite{2000_ppd}.

Magnon condensation in the lowest energy level(s) occurs when the system-specific critical magnon density $n_{\rm c}$ is exceeded. Roughly speaking, $n_{\rm c}$ corresponds to the point when the inter-magnon separation becomes comparable to the thermal de Broglie wavelength $\lambda_{\rm dB}$, i.e. when $n_{\rm c}^{-1/3} \sim \lambda_{\rm dB}$. At this point the chemical potential of the magnon system $\mu$ asymptotically approaches the ground state energy $\epsilon_0$ and the ground state population
\begin{equation}
    n_0 = \frac{1}{e^{(\epsilon_0 - \mu)/k_{\rm B}T}-1}
\end{equation}
diverges. Here $k_{\rm B}$ is the Boltzmann constant and $T$ is temperature. As a result, the ground state becomes populated by a macroscopic number of constituent particles that spontaneously form a macroscopically coherent state that is stable against decohering perturbations.

Bose-Einstein condensates of non-conserved quasiparticles are ubiquitous in nature; similar phenomenology is used to describe systems consisting of phonons \cite{MISOCHKO2004381}, rotons \cite{PhysRevB.84.024525}, photons \cite{PhotonBEC}, excitons \cite{ExcitonBEC}, exciton-polaritons \cite{RevModPhys.82.1489}, etc. To date BECs consisting of magnons, in particular, have been reported in various superfluid phases of $^3$He \cite{1984_hpd_exp,PhysRevLett.121.025303,Hunger_3HeABEC}, in cold atomic gases \cite{PhysRevLett.114.125304,PhysRevLett.116.095301}, and in a few solid-state systems \cite{Demokritov2006, PhysRevLett.108.177002}. Moreover, the antiferromagnetic hematite ($\alpha$-Fe$_2$O$_3$) has been put forwards as a promising candidate system for condensation of magnons \cite{Bunkov_2018, 9706176}. We note that while the concept of magnon BEC is also useful for describing the onset of magnetic-field-induced magnetic order in spin-dimer compounds \cite{PhysRevLett.84.5868}, such as TlCuCl$_3$, the excitations in such systems are in thermal equilibrium and therefore the chemical potential is always zero, i.e. $\mu = 0$. While the phenomenology is similar \cite{MagneticBEC}, such systems are outside the scope of this Perspective.

The nature and experimental realization of the magnon BEC varies widely between different physical systems. In $^1$H, a magnon BEC was created at high magnetic field of multiple tesla by initially preparing a dense cloud of cold gas in a higher-energy low-field-seeking spin state, while magnons are created by pumping a number of atoms to a lower-energy, high-field-seeking state \cite{PhysRevLett.114.125304}. The magnons become collective excitations via an effect known as "identical spin rotation", where a single spin-flip is carried through multiple atom-atom collisions \cite{ISR1,ISR2,ISR3}. In the $^{87}$Rb $F=1$ spinor condensate the magnon (quasi-) BEC results from spin-exchange collisions between different internal spin states in the same hyperfine manifold at low magnetic fields \cite{PhysRevLett.116.095301}. In superfluid phases of $^3$He, magnons are collective excitations stemming from the Nambu-Goldstone modes of the underlying order parameter \cite{vollhardt2013superfluid}. Finally, out of the solid state systems where magnon BEC has been realized, perhaps the most promising systems are thin Yttrium-iron-garnet (YIG) films, where the magnon condensate is created either by continuous radio frequency pumping \cite{Demokritov2006} or by laser-induced spin currents \cite{YIGSpinCurrMagnonBEC} at room temperature in highly non-equilibrium state and, moreover, in the momentum space instead of in real space as for the other examples.

Regardless of the system, a magnon BEC can be viewed as the ground state of the relevant subsystem, which is an essential viewpoint in the context of time crystals. For example, if one ignores the weak non-conservation of magnons in $^3$He-B caused by the spin-orbit interaction, the ground state of the subsystem would evolve in time, realizing the time crystal as originally suggested by Wilczek \cite{PhysRevLett.109.160401}. However, due to conservation of particles the time evolution of such a system is unobservable \cite{PhysRevLett.110.118901,PhysRevLett.111.070402,Nozieres_2013,Volovik2013}. A similar scenario is realized in atomic BECs as well as in superconductors (in superconductors the coherent precession is that of Anderson pseudospins), where the evolution of the superfluid phase is unobservable. In principle, for long enough measurement times, the phase should become observable due to proton decay. In some cases, such as in $^1$H and in $^3$He, the absolute value of the phase can be directly monitored in real time as the phase of the precessing magnetization can be measured by oriented pickup coils which provide the necessary loss channel. Such direct measurement of the absolute value of the phase of the macroscopic wave function is rather uncommon and can be exploited for a variety of purposes, as discussed later in this Perspective.

The spontaneously formed coherent precession of magnetization has many faces: spin superfluidity, off-diagonal long-range order (ODLRO), the Bose-Einstein condensation of nonequilibrium (pumped) quasiparticles, and time crystal. In Section~\ref{MagnonBEC}, we briefly describe the basic properties of the magnon BEC. Section~\ref{TimeCrystal} focuses on magnon-BEC time crystals. Magnon BECs can be utilized to simulate different processes and objects in particle physics, such as spherical charge solitons (Section~\ref{LocalizationSec}), the light Higgs particle (Section~\ref{Higgs}) and analogue event horizons (Section~\ref{sec:spacetime}).  Section~\ref{sec:probe} concentrates on probing topological defects using magnon BEC and, finally, Section~\ref{sec:outlook} contains an outlook on future prospects of magnon BECs in various systems.

\begin{figure}[htb!]
\includegraphics[width=0.7\linewidth]{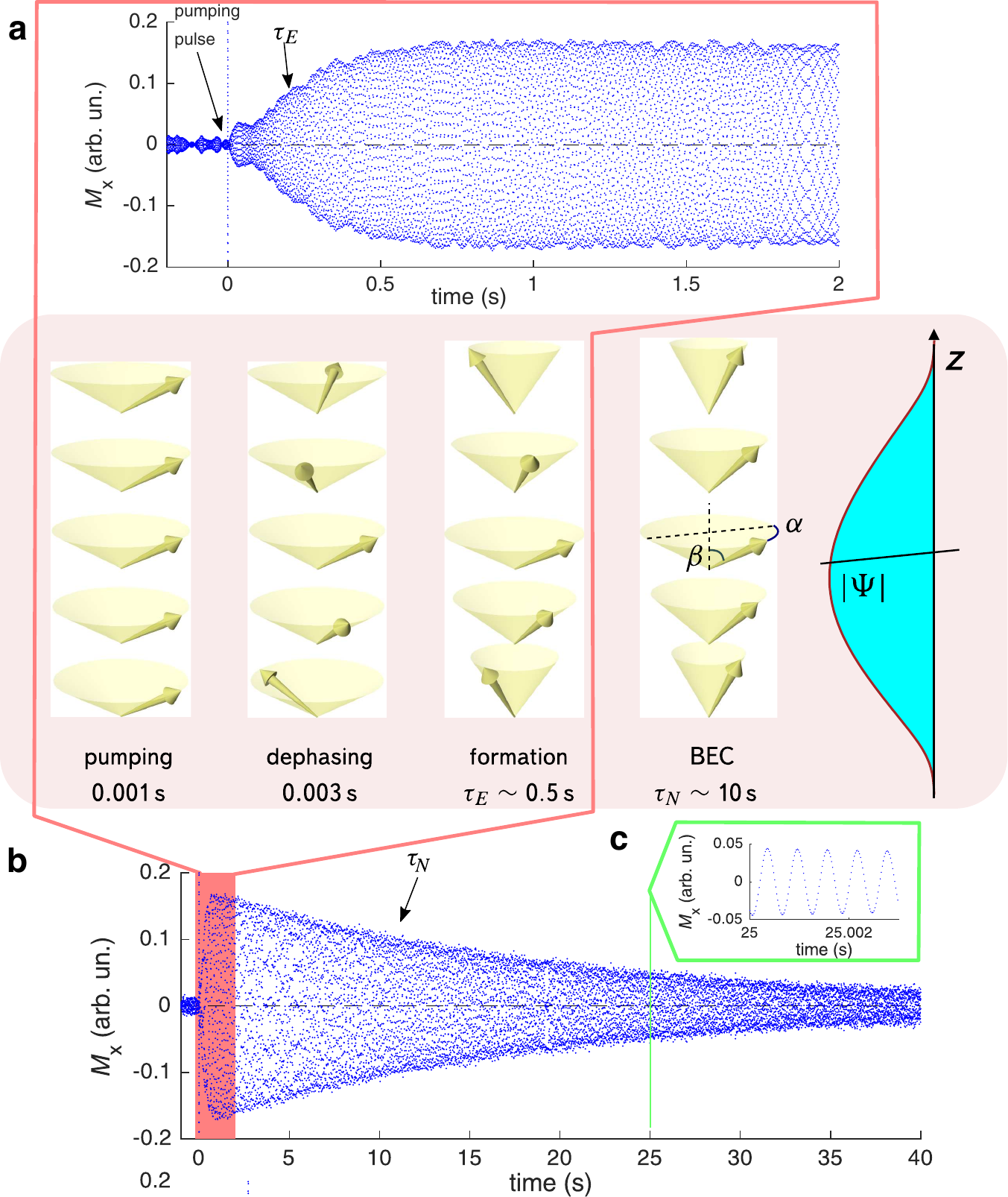}
\caption{{\bf Observing magnon Bose-Einstein condensation:} ({\bf a}) Magnons are pumped to the system with a radio-frequency pulse at zero time, seen as the sharp peak in the data. As illustrated in the central panel on a colored background, the pumping is followed by dephasing of the precession. If the magnon density is high enough, a BEC emerges after $\tau_{\rm E}$, manifest in coherent precession of magnetization $M_x+iM_y\propto\left<\hat S^+\right>= \sqrt{2S}\left<\hat a_0\right>=S_\perp e^{i\omega t}$. This is picked up by the NMR coils and measured as an oscillating voltage. Magnetic relaxation in superfluid $^3$He is very slow, and the number of magnons $N$ decreases with time constant $\tau_{\rm N}$ (here $\tau_{\rm N} \sim 10$\,s), seen as a slow decrease in the signal amplitude, shown in panel ({\bf b}). Panel ({\bf c}) shows a further zoom-in into the band indicated by the green line. Here, the sinusoidal pick-up signal generated by the precession of magnetization is clearly seen. Data shown in this figure was measured at 0\,bar pressure and \SI{131}{\micro\kelvin} temperature \cite{PhysRevLett.120.215301}.
}
\label{Precession}
\end{figure}

\section{Coherent precession and spin superfluidity}
\label{MagnonBEC}

\subsection{Off-diagonal long-range order}

The phenomenon of Bose-Einstein condensation was originally suggested by Einstein for stable particles with integer spin. Under suitable conditions, the BEC gives rise to macroscopic phase coherence and superfluidity, first observed in liquid $^4$He. This a consequence of the spontaneous breaking of the global $U(1)$ gauge symmetry related to the conservation of the particle number $N$, e.g. of $^4$He atoms.

As distinct from many other systems with spontaneously broken symmetry, such as crystals, liquid crystals, ferro- and antiferromagnets, the order parameter in superfluids and superconductors is manifested in the form known as the off-diagonal longe-range order (ODLRO). In bosonic superfluids (such as liquid  $^4$He) the manifestation of the ODLRO is that the average values of the creation and annihilation operators for the particle number are nonzero in the superfluid state, i.e., 
\begin{equation}
\Psi = \left<\hat{\mbox{\boldmath$\Psi$}}\right> \,\,,
\,\, \Psi^* = \left<\hat{\mbox{\boldmath$\Psi$}}^\dagger\right>
\,.
 \label{ODLROsuper}
\end{equation}%
In conventional (i.e. not superfluid or superconducting) states the creation or annihilation operators have only the off-diagonal matrix elements,  such as $\left<N|\hat{\mbox{\boldmath$\Psi$}}^\dagger|N+1\right>$, describing the transitions between states with different number of particles. In the thermodynamic limit $N \rightarrow \infty$, the states with different numbers of particles in the Bose condensate are not distinguished, and the creation or annihilation operators acquire the nonzero average values. In superconductors and fermionic superfluids such as superfluid $^3$He, the ODLRO is represented by the average value of the product of two creation  $\left<a^\dagger_{\bf k}a^\dagger_{-{\bf k}}\right>$ or two annihilation  $\left<a_{\bf k}a_{-{\bf k}}\right>$  operators, which reflects the Cooper pairing in fermionic systems. In quantum theory, states with nonzero values of the creation or annihilation operators are called squeezed coherent states.

\subsection{ODLRO and coherent precession}

The magnetic ODLRO  can be represented in terms of magnon condensation, applying the Holstein-Primakoff transformation. The spin operators are expressed in terms of the magnon creation and annihilation operators
\begin{eqnarray}
\hat a_0~\sqrt{1-\frac{\hbar a^\dagger_0 a_0}{2{\mathcal S}}}= \frac{\hat {\mathcal S}^+}{\sqrt{2{\mathcal S}\hbar}} \,\,,\,\, \sqrt{1-\frac{\hbar a^\dagger_0 a_0}{2{\mathcal S}}}~\hat a^\dagger_0=
\frac{\hat {\mathcal S}^-}{\sqrt{2{\mathcal S}\hbar}}~,
\label{MagnonCreation}
\\
\hat {\mathcal N}=\hat a^\dagger_0\hat a_0 =
\frac{{\mathcal S}-\hat{\mathcal S}_z}{\hbar}~.
\label{MagnonNumberOperator}
\end{eqnarray}

Eq. (\ref{MagnonNumberOperator}) relates the number of magnons ${\mathcal N}$ to the deviation of spin ${\mathcal S}_z$ from its equilibrium value ${\mathcal S}_z^{({\rm equilibrium})}={\mathcal S}=\chi H V/\gamma$, where $\chi$ and $\gamma$ are spin susceptibility and gyro-magnetic ratio, respectively. Pumping ${\mathcal N}$ magnons into the system (e.g. by a RF pulse) reduces the total spin projection by $\hbar {\cal N}$, i.e. ${\mathcal S}_z ={\mathcal S} - \hbar {\mathcal N}$. The ODLRO in magnon BEC is given by:
\begin{equation}
\left<\hat a_0 \right>={\mathcal N}^{1/2}  e^{i\omega
t+i\alpha}=\sqrt{\frac {2{\mathcal S}}{\hbar}} ~\sin\frac{\beta}{2} ~e^{i\omega
t+i\alpha}\,,
\label{ODLROmagnon}
\end{equation}
where $\beta$ is the tipping angle of precession. The role of the chemical potential $\mu$ is played by the global frequency of the coherent precession $\omega$, i.e. $\mu \equiv \hbar \omega$ and the phase of precession $\alpha$ plays the role of the phase of the condensate, i.e. $\Phi\equiv \alpha$. A typical experimental signal showing an exciting pulse, the formation of the BEC and the slow decay is shown and analysed in Fig.~\ref{Precession}. The experimental setup used in this particular experiment is shown in Fig.~\ref{Fig:ExpSetup}. Note that the analogy with atomic BECs is valid only for the dynamic states of the magnetic subsystem and not e.g. for static magnets with $\omega=0$.

\subsection{Gross-Pitaevskii and Ginzburg-Landau description}
\label{sec:GP}

As for atomic Bose condensates, the magnon BEC is described by the Gross-Pitaevskii equation.
The local order parameter is obtained by extension of Eq.~(\ref{ODLROmagnon}) to the inhomogeneous case, $\hat a_0 \to \hat\Psi({\bf r},t)$, and is determined as the vacuum expectation value of the magnon field operator:
\begin{equation}
\Psi({\bf r},t)=\left<\hat \Psi({\bf r},t)\right>~~,~~ n=\vert\Psi\vert^2~~,~~{\mathcal N}=\int d^3r ~\vert\Psi\vert^2~.\label{OrderParameter1}
\end{equation}
where $n$ is the magnon density.

If the dissipation and pumping of magnons are ignored, the corresponding Gross-Pitaevskii equation has the conventional form:
  \begin{equation}
-i \hbar \frac{\partial \Psi}{\partial t}= \frac{\delta {\mathcal F}}{\delta \Psi^*}~,
\label{GP}
\end{equation}
where ${\mathcal F}\{ \Psi\}$ is the free energy functional forming the effective Hamiltonian of the spin subsystem. In the coherent precession, the global frequency is constant in space and time
\begin{equation}
\Psi({\bf r},t)=\Psi({\bf r})  e^{i\omega t}~,
\label{OrderParameter2}
\end{equation}
and the Gross-Pitaevskii equation transforms into the Ginzburg-Landau equation with $\hbar \omega=\mu$:
  \begin{equation}
 \frac{\delta {\mathcal F}}{\delta \Psi^*}- \mu\Psi=0~.
\label{GL}
\end{equation}

\begin{figure}
\includegraphics[width=0.6\linewidth]{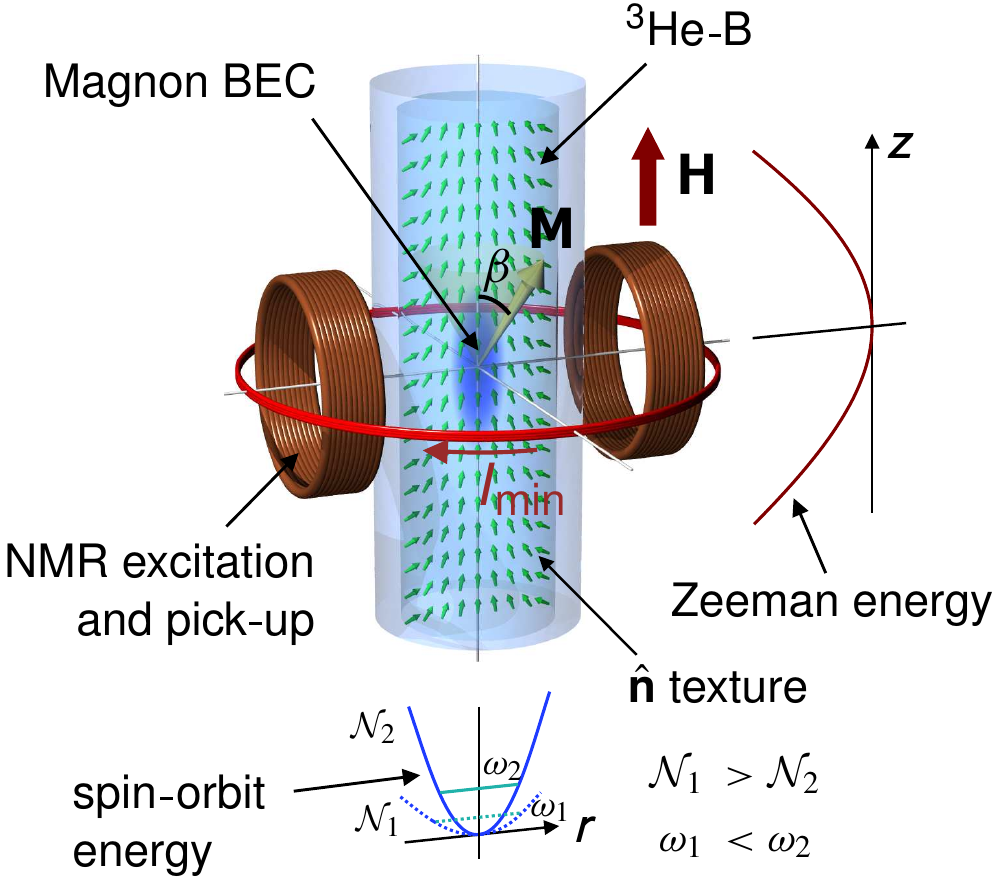}
\caption{{\bf Magnon BEC in magneto-textural trap in superfluid $^3$He.} The magnetization $\mathbf{M}$ of the condensate is deflected by an angle $\beta$ from the direction of magnetic field $\mathbf{H}$ and precesses coherently around the field direction with the frequency $\omega$. Magnons are confined to the nearly harmonic 3-dimensional trap formed by the spatial variation of the field $\mathbf{H}(\mathbf{r})$ via Zeeman energy and by the spatial variation (texture) of the orbital anisotropy vector $\hat{\mathbf{n}}(\mathbf{r})$ of $^3$He-B via spin-orbit interaction energy. The orbital texture is flexible and yields with increasing number of magnons ${\cal N}$ in the trap, resulting in lower radial trapping frequency. As a result, the chemical potential, observed as the precession frequency, decreases. This leads to inter-magnon interaction (Sec.~\ref{sec:GP}) and eventually to Q-ball formation (Sec.~\ref{Qball}).
}
\label{Fig:ExpSetup}
\end{figure}

The free energy functional reads
  \begin{eqnarray}
  {\mathcal F} -\mu {\mathcal N}=\int d^3r
   \left(
\frac{1}{2} g^{ik}\nabla_i \Psi^* \nabla_k\Psi  + \hbar (\omega_{\rm L}({\bf r})-\omega)
  \vert\Psi\vert^2 +F_{\rm so}(\vert\Psi\vert^2)\right) \,,
  \label{GLfunctional}
\end{eqnarray}
where $\omega_{\rm L}$ is the local Larmor frequency $\omega_{\rm L}({\bf r})=\gamma H({\bf r})$ and $g^{ik}$ describes rigidity of the magnon system. The spin-orbit interaction energy $F_{\rm so}$ is a sum of contributions proportional to $\vert\Psi\vert^2$ and $\vert\Psi\vert^4$, see e.g.\ Ref.~\citenum{bunkov2012spin}. Thus, the free energy functional can be compared with the conventional Ginzburg-Landau free energy of an atomic BEC:
 \begin{eqnarray}
  {\mathcal F} -\mu {\mathcal N}=\int d^3r
   \left(
\frac{1}{2} g^{ik}\nabla_i \Psi^* \nabla_k\Psi +(U({\bf r})-\mu)
  \vert\Psi\vert^2 +b\vert\Psi\vert^4\right)
  \label{GLfunctionalBEC}
\end{eqnarray}
with the external potential $U({\bf r})$ formed by the magnetic field profile and a part of the spin-orbit interaction energy. The fourth order term, which describes the interaction between magnons, originates from the rest of spin-orbit interaction $F_{\rm so}(\vert\Psi\vert^2)$. Fig.~\ref{Fig:ExpSetup} illustrates the appearance of the inter-magnon interaction via flexible orbital texture in the case of trapped magnon BEC.

The gradient energy in Eq.~(\ref{GLfunctionalBEC}) is responsible for establishing coherence across the sample. In the London limit, it can be expressed via gradients of the precession phase $\alpha$
\begin{equation}
F_{\rm grad}= \frac{1}{2} g^{ik}\nabla_i \Psi^* \nabla_k\Psi =
\frac{1}{2}K_{ik} \nabla_i \alpha \nabla_k\alpha =
\frac{1}{2} n \left(m^{-1}\right)_{ik}  \nabla_i \alpha  \nabla_k \alpha
\,.
 \label{rigidity}
\end{equation}
A necessary condition for spin superfluidity and phase coherence is that the gradient energy is positively determined. This condition is not universally valid in all systems with magnons, but is applicable, e.g., in $^3$He-B. The spin superfluid currents are then generated by the gradient of the phase. The rigidity tensor $K_{ik}$ can be further expressed via magnon mass $m_{ik}(n)$, which in general depends on the magnon density $n$ and is anisotropic due to applied magnetic field \cite{Zavjalov_2015}.

\section{Magnon BEC as a time crystal}
\label{TimeCrystal}


Originally, time crystals were suggested as class a quantum systems for which time translation symmetry is spontaneously broken in the ground state \cite{PhysRevLett.109.160401}. It was quickly pointed out that the original concept cannot be realized and observed in experiments, essentially because that would constitute a perpetual motion machine \cite{PhysRevLett.110.118901,PhysRevLett.111.070402,Nozieres_2013,Volovik2013}. That is, if the system is strictly isolated, i.e. when the number of particles $N$ is conserved, there is no reference frame for detecting the time dependence \cite{PhysRevResearch.3.L042023}. This no-go theorem led researchers to search for spontaneous breaking of the time-translation symmetry on more general grounds, turning to out-of-equilibrium phases of matter (see e.g. reviews ~\citenum{Sacha_2017,hannaford2022decade,sacha2020discrete}). With this adjustment, feasible candidates of time-crystal systems include those with off-diagonal long range order, such as superfluids \cite{Annett:730995}, Bose gases \cite{PhysRevLett.121.185301}, and magnon condensates \cite{Volovik2013}. 

Any system with ODLRO can be characterized by two relaxation times \cite{Volovik2013}: the lifetime of the corresponding (quasi)particles $\tau_{\rm N}$ and the thermalization time $\tau_{\rm E}$ during which the BEC is formed. If $\tau_{\rm N} \gg \tau_{\rm E}$, the system has sufficient time to relax to a minimal energy state with (quasi-)fixed ${\cal N}$ (i.e. to form the condensate). During the intermediate interval $\tau_{\rm N} \gg t \gg \tau_{\rm E}$ the system has finite $\mu$ corresponding to spontaneously-formed uniform precession that can be directly observed as shown in Fig.~\ref{Precession}. In $^3$He-B $\tau_{\rm N}$ can reach tens of minutes at the lowest temperatures \cite{2000_ppd} -- this is the closest an experiment has got to a time crystal in equilibrium. 

Finally, we point out that in the grand unification theory extensions of Standard Model the conservation of the number of atoms is absent due to proton decay \cite{NATH2007191}. Therefore, in principle, the oscillations of an atomic superfluid in its ground state can be measured, albeit the time scale for the decay is at least in the $\sim 10^{36}\,$ years range \cite{NATH2007191}.

\subsection{Discrete and continuous magnon time crystals}
\label{TimeQuasi}

Time crystals are commonly divided into two broad categories based on their symmetry classification. If, relative to the Hamiltonian before the phase transition to the time crystal phase, the system spontaneously breaks the discrete time translation symmetry, it is called a {\it discrete time crystal}. Such a system is realized e.g. in parametric pumping scenarios, when the periodicity of the formed time crystal differs from that of the drive. On the other hand, if the spontaneously broken symmetry is the {\it continuous} time translation symmetry (the preceding Hamiltonian is not periodic in time), the system is a {\it continuous time crystal}. Note that both discrete and continuous time crystals may still possess a discrete time translation symmetry.

Both types of time crystals have been observed in magnon BEC experiments \cite{PhysRevLett.120.215301,autti2020,autti2022nonlinear, doi:10.1126/science.abo3382, doi:10.1146/annurev-conmatphys-031119-050658}. Discrete time crystals are realized under an applied RF drive, when the frequency of the coherent spin precession deviates from that of the drive. If the induced precession frequency is incommensurate with the drive, the system obtains the characteristics of a discrete time quasicrystal. On the other hand, if the magnon number decay is sufficiently slow, i.e., $\tau_{\rm N}^{-1} \ll \omega$, where $\omega$ is the frequency of motion of the time crystal, the coherent precession can be observed for a long time after the drive has been turned off. This is a continuous time crystal, which can generally be formed in magnon BEC materials as $\tau_{\rm E}^{-1} < \omega$ by definition. In $^3$He-B continuous time crystals reach life times longer than $10^{7}$ periods \cite{PhysRevLett.120.215301}. 

\subsection{Phonon in a time crystal}
\label{sec:PhononsInTC}

\begin{figure}[t!]
\centerline{\includegraphics[width=0.5\linewidth]{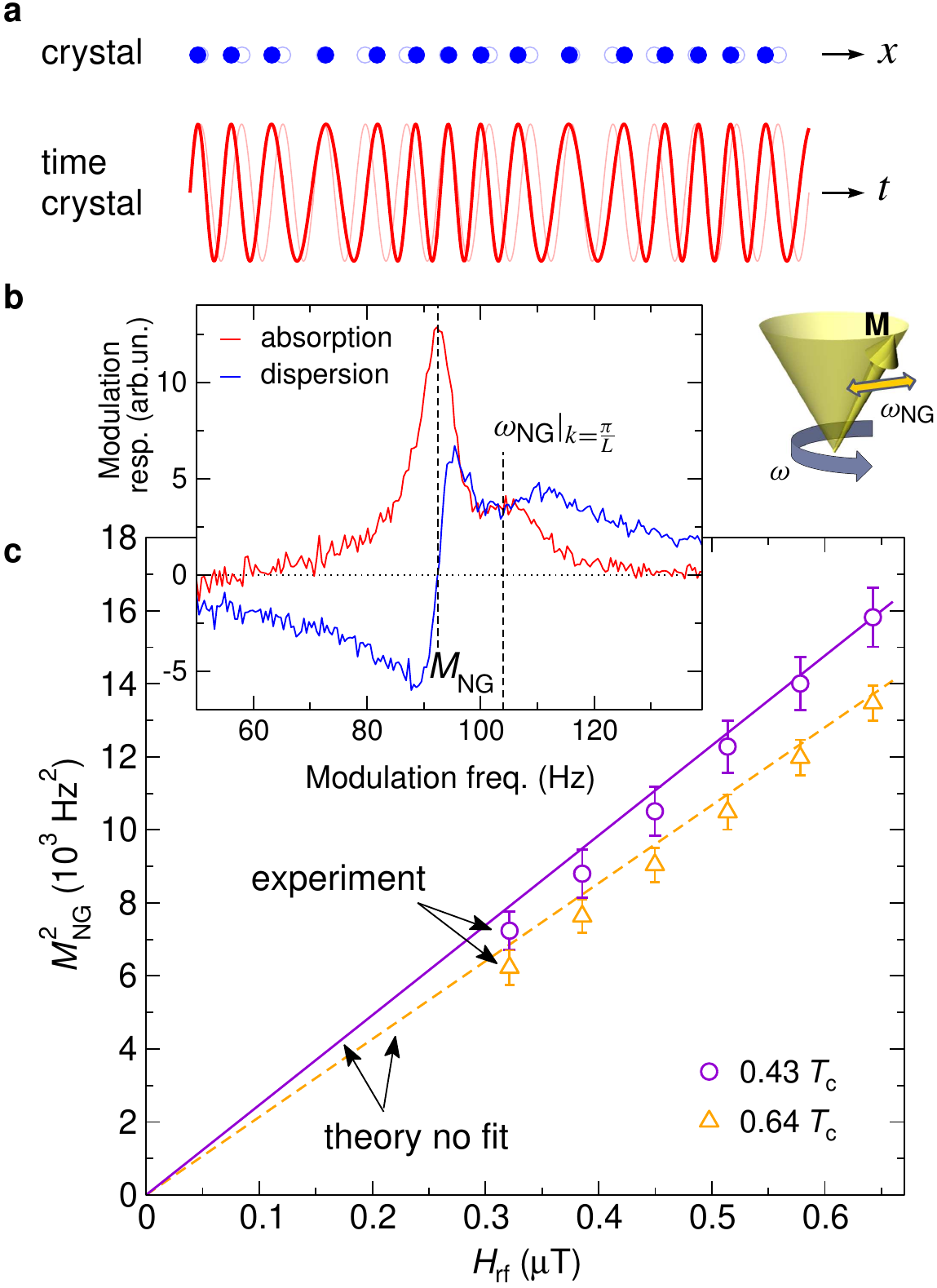}}
\caption{\textbf{Phonon in a magnon-BEC time crystal.} (\textbf{a}) In a crystal in a ground state, atoms occupy periodic locations in space (empty circles), while phonon excitation results in a periodic shift from these positions (filled circles). A time crystal is manifested by a periodic process (thin line), and a phonon excitation leads to periodic variation of the phase of that process (thick line). (\textbf{b}) In a magnon-BEC time crystal, the periodic process is the precession of magnetization $\mathbf{M}$ at frequency $\omega$. The phonon excitation modulates the precession phase at the frequency $\omega_{\rm NG}$ (right). This mode can be excited by applying modulation to the rf drive of the condensate and observed by detecting response in the induction signal from the pick-up coil at the modulation frequency (left). Two standing-wave modes in the sample are visible (marked with the vertical dashed lines). (\textbf{c}) As the measurements on the panel (b) are done with finite rf excitation of the amplitude $H_{\rm rf}$, the phonon becomes a pseudo-Nambu-Goldstone mode with the mass $M_{\rm NG}$, Eq.~(\ref{eq:pseudoNG}). Extrapolation to the freely evolving time crystal at zero $H_{\rm rf}$ shows that the phonon becomes massless, as expected for spontaneous symmetry breaking. The measurements are done at two temperatures at a pressure of 7.1\,bar in the polar phase of $^3$He \cite{PhysRevLett.121.025303}.
\label{fig:ngmode}}
\end{figure}

Spontaneous breaking of continuous time translation symmetry in a regular crystal results in the appearance of the well-known Nambu-Goldstone mode -- a phonon. Similarly, the spontaneous breaking of time translation symmetry in a continuous time crystal should lead to a Nambu-Goldstone mode, manifesting itself as an oscillation of the phase of the periodic motion of the time crystal, Fig.~\ref{fig:ngmode}a. This mode can be called a phonon in the time crystal.

In time crystals formed by magnon BECs, the phononic mode is equivalent to the Nambu-Goldstone mode related to spin-superfluid phase transition\cite{PhononsInTC}. It is easier to excite in experiments when the spin precession is driven by a small applied rf field $H_{\rm rf}$ by modulating the phase of the drive, Fig.~\ref{fig:ngmode}b. In this case the time translation symmetry is already broken explicitly by the drive, and the phonon becomes a pseudo-Nambu-Goldstone mode with the mass (gap) $M_{\rm NG}$. Its dispersion relation connecting the wave vector $k$ and the frequency $\omega_{\rm NG}$ becomes 
\begin{equation} \label{eq:pseudoNG}
    \omega_{\rm NG}^2 = M_{\rm NG}^2 + c_{\rm NG}^2 k^2\,.
\end{equation}
For the sample size of $L$, standing-wave resonances can be seen for $k = n \pi/L$, where $n$ is an integer, Fig.~\ref{fig:ngmode}b, and the mass $M_{\rm NG}$ and the propagation speed $c_{\rm NG}$ can be determined. According to the theory $M_{\rm NG} \propto H_{\rm rf}$ -- experiments in the polar phase of $^3$He demonstrate excellent agreement without fitting parameters \cite{PhysRevLett.121.025303}, Fig.~\ref{fig:ngmode}c. This mode was also observed in time crystals formed by magnon BEC in the B phase of $^3$He \cite{PhysRevLett.100.155301, Goldstone3HeB}. Extrapolation of the mass to the case of a freely evolving time crystal at $H_{\rm rf} =0$ leads to a true massless Nambu-Goldstone mode -- a phonon in a time crystal.

\subsection{Interacting time crystals}
\label{Josephson}

Interacting time crystals have been realized in $^3$He-B by creating two continuous time crystals with close natural frequencies close to each other \cite{autti2022nonlinear, autti2020}. In a magneto-textural trap such as used in Refs.~\citenum{autti2022nonlinear, autti2020, PhysRevLett.120.215301} the radial trapping potential is provided by the spin-orbit interaction via the spatial order parameter distribution. The magnetic feedback of the magnon number to the order parameter texture means the time crystal frequency (period) is regulated internally, $\omega_{\rm B} ({\cal N}_{\rm B})$.\cite{Autti2012} The frequency increases as the magnon number slowly decreases (the decay mechanism is not important but details can be found in Refs.~\citenum{magnon_relax,Heikkinen2014}). The second time crystal is created against the edge of the superfluid where such feedback is suppressed. The result is a macroscopic two-level system described by the Hamiltonian
\begin{equation}
    {\cal H} = \hbar
    \begin{pmatrix}
     \omega_{\rm B}[{\cal N}_{\rm B}(t)] & -\Omega \\
     -\Omega & \omega_{\rm S}
    \end{pmatrix}\,
\end{equation}
where the coupling $\Omega$ is determined by the spatial overlap of the time crystals' wave functions.

\begin{figure}
\includegraphics[width=0.7\linewidth]{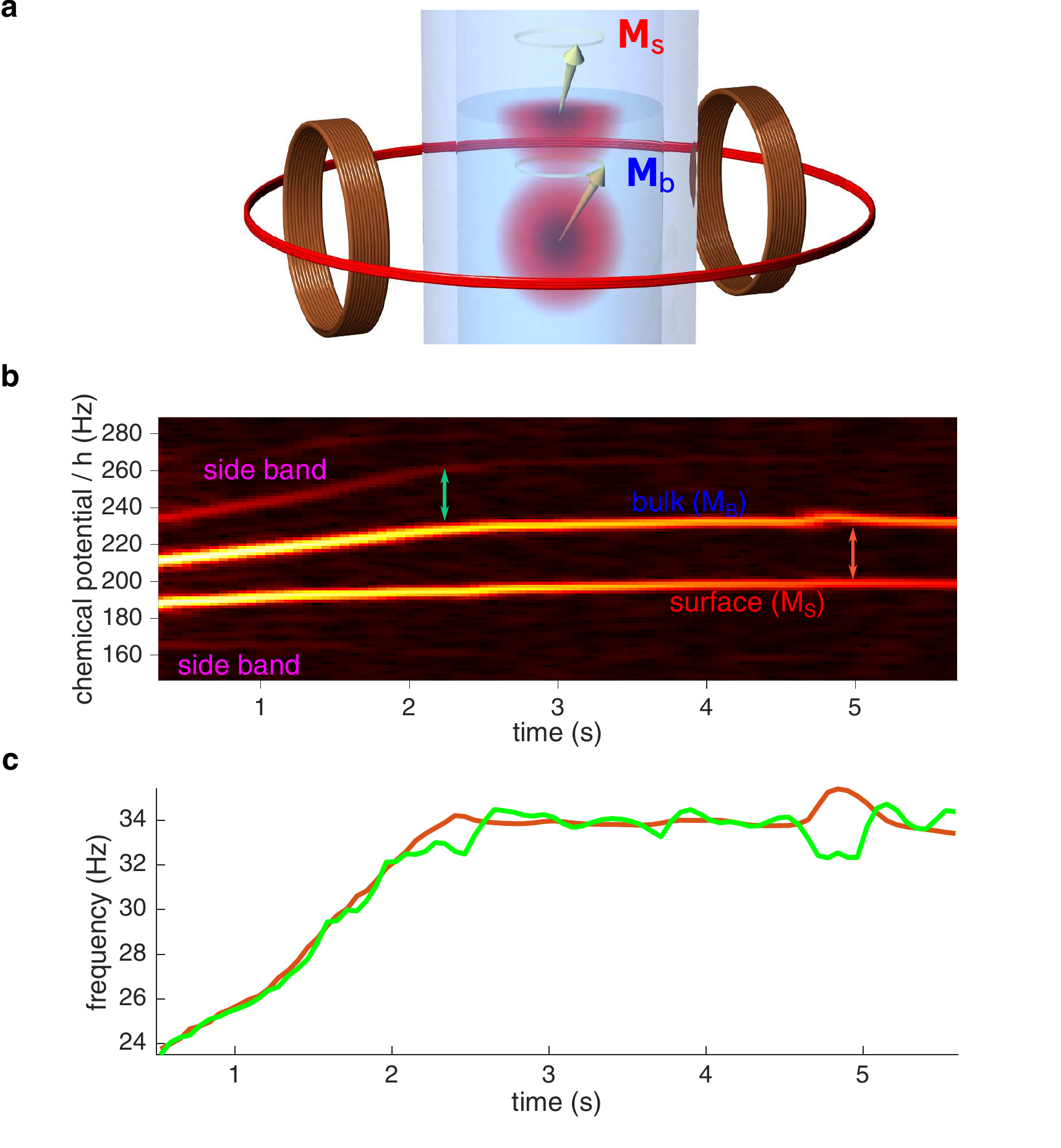}
\caption{{\bf Josephson (Rabi)  population oscillations between two magnon-BEC time crystals.} ({\bf a}) We can create two local minima for magnon-BEC time crystals, one in the bulk and one against a free surface of the superfluid. ({\bf b}) Both are populated with a pulse at zero time, after which the bulk frequency is slowly changing due to changes in the trap shape as the magnon number slowly decreases. The two levels are coupled, resulting in Josephson population oscillations between them, observed as the side bands above and below the main traces. The side band frequency separation (green arrow), shown by the green line in panel {\bf c}, corresponds to the separation of the main traces (red arrow), shown by the red line in panel {\bf c}. This ties the population oscillation to the chemical potential difference of the time crystals and thus to Josephson oscillations. The oscillations the bulk population and the surface population are shown to take place with opposite phases in Ref.~\citenum{autti2020}. The Josephson oscillations become Rabi oscillations if the two-level frequencies are brought close to one another as explained in Ref.~\citenum{autti2022nonlinear}.}
\label{Fig:interactingTC}
\end{figure}

In this configuration, the two time crystals may interact by exchanging the constituent quasiparticles. The exchange of magnons results in opposite-phase oscillations in the respective magnon populations of the two time crystals (Fig.~\ref{Fig:interactingTC}) which is equivalent to the AC Josephson effect in spin superfluids \cite{30002064863}. 

\begin{figure}
\includegraphics[width=0.6\linewidth]{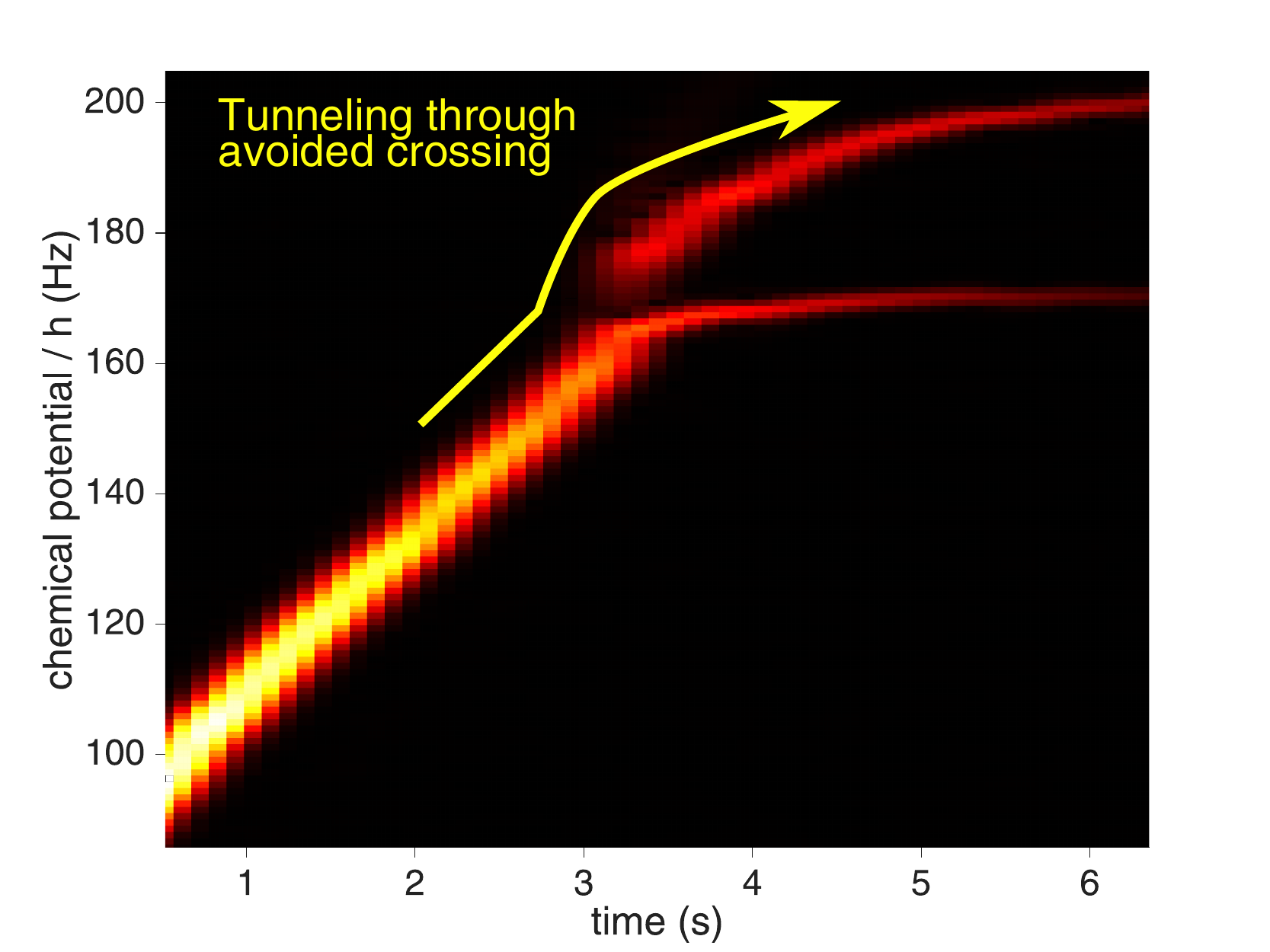}
\caption{{\bf Landau-Zener tunneling between two magnon-BEC time crystals.} One of the two levels is populated at time zero with an exciting pulse (framed out to emphasise the rest of the signal). The chemical potential of this state increases gradually as the magnon number decays, crossing the second level after 3\,s. As the avoided crossing is traversed at finite rate (not adiabatically), a part of the magnon population tunnels to the excited level at the avoided crossing \cite{autti2022nonlinear}.   }
\label{Fig:LZ-simple}
\end{figure}

Further two-level quantum mechanics can be accessed by making the precession frequencies of two time crystals cross during the experiment using the dependence $\omega_{\rm B} ({\cal N}_{\rm B})$. The result is magnons moving from the ground state to the excited state of the two-level Hamiltonian in a Landau-Zener transition, see Fig.~\ref{Fig:LZ-simple}. Remarkably, these phenomena are directly observable in a single experimental run, including the chemical potentials and absolute phases of the two time crystals, implying that such basic quantum mechanical processes are also technologically accessible for magnonics and related quantum devices.

\section{Magnon BEC and cosmology}
\label{LocalizationSec}

In the context of particle physics and cosmology, the magnon BEC provides a laboratory test bench for otherwise inaccessible or convoluted theoretical concepts. This phenomenology may eventually play a major role in the technological toolbox for magnonics, albeit potential applications cannot yet be predicted. Here we will discuss the analogs between trapped magnons and two cosmological concepts: the $Q$-ball and the MIT bag model.

\subsection{Magnonic $Q$-ball}
\label{Qball}

Self-bound macroscopic objects encountered in everyday life are made of fermionic matter, while bosons mediate interactions between and within them. Compact (self-bound) objects made purely from interacting bosons may, however, be stabilized in relativistic quantum field theory by conservation of an additive quantum number $Q$ \cite{coleman_qball,Cohen1986301,friedberg_qball}. Spherically symmetric non-topological $Q$-charge solitons are called $Q$-balls. They generally arise in charge-conserving relativistic scalar field theories. 

Observing $Q$-balls in the Universe would have striking consequences beyond supporting supersymmetric extensions of the Standard Model \cite{Kusenko199846,Enq} -- they are a candidate for dark matter \cite{Kasuya,Kus2,Enq,Kus}, may play a role in the baryogenesis \cite{Kasuya2} and in formation of boson stars \cite{BosonStarPRD.77.044036}, and supermassive compact objects in galaxy centers may consist of $Q$-balls \cite{Troitsky2016}. Nevertheless, unambiguous experimental evidence of $Q$-balls has so far not been found in cosmology or in high-energy physics. Analogues of $Q$-balls have been speculated to appear in atomic BECs in elongated harmonic traps \cite{EnqvistLaine} and possibly play a role in the $^3$He A-B transition puzzle \cite{Hong_3HeAQBall}. Additionally, the properties of bright solitons in 1D atomic BECs \cite{1D_lithium_Qball} and Pekar polarons in ionic crystals \cite{DevreesePolaron} bear similarities with $Q$-balls.

\begin{figure}
\includegraphics[width=0.7\linewidth]{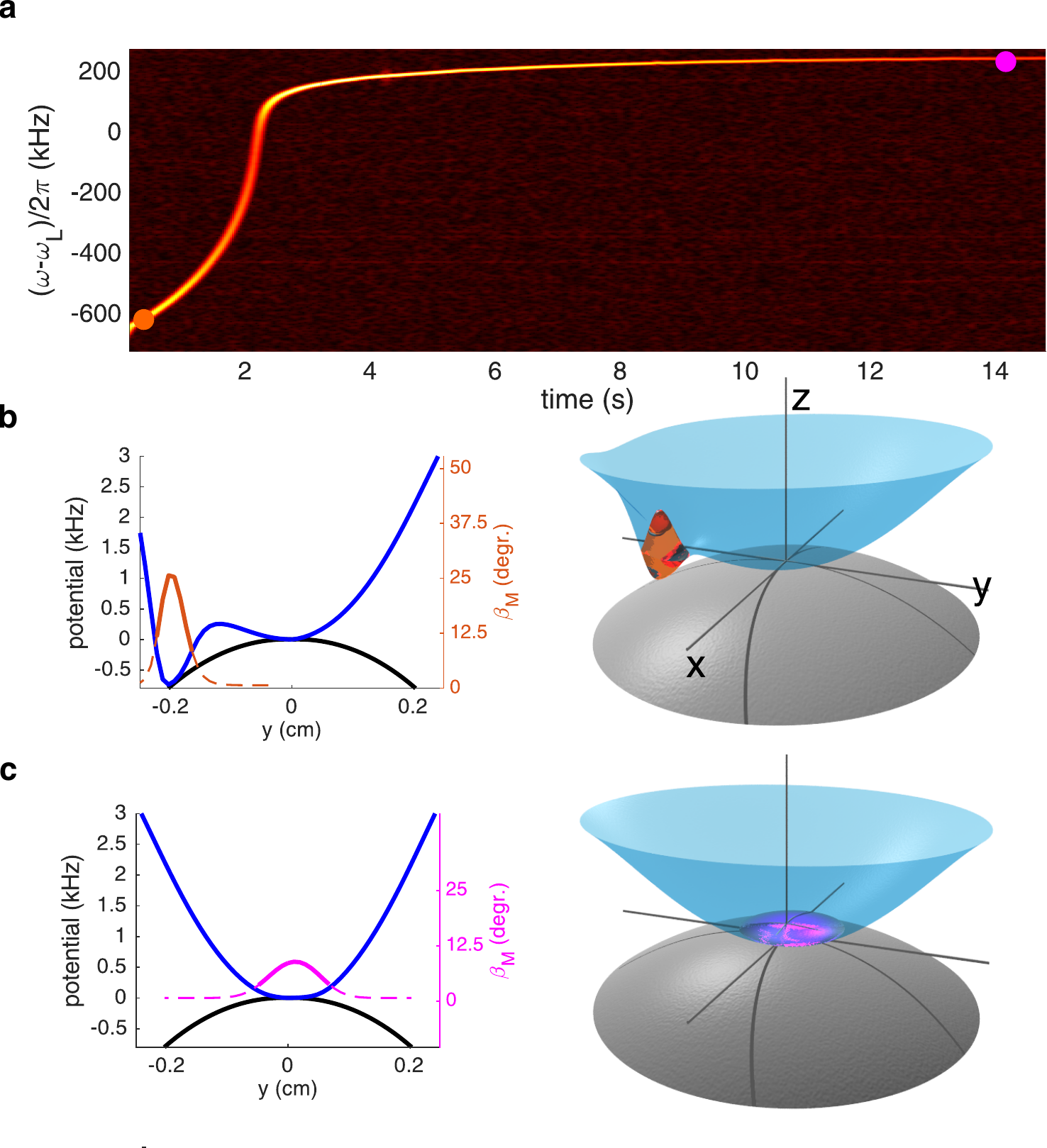}
\caption{{\bf Magnon BEC as a self-propagating Q-ball soliton.} ({\bf a}) The magnon BEC is created with a pulse (not shown) with a very large initial magnon number. The time-dependent frequency spectrum of the recorded signal is shown here in such a way that dark corresponds to no magnons and yellow to large magnon population. In the beginning (red dot and panel {\bf b}) the BEC (orange line and orange blob) is located in a side minimum, where the trapping potential (blue line, blue surface) is modified down to the unyielding trap component controlled by the magnetic field (black line, dark surface). As the magnon number decreases due to slow dissipation, the trapping potential evolves and the BEC gradually moves across several millimeters to the symmetric central position (magenta dot and panel ({\bf c})). Here the BEC is illustrated by the green line and blob. \cite{autti2018}
}
\label{Fig:qball-large}
\end{figure}

The trapped magnon BEC in $^3$He-B provides a one-to-one implementation of the $Q$-ball Hamiltonian. The charge $Q$ is the number of magnons and the BEC precession frequency corresponds to the frequency of oscillations of the relativistic field within the $Q$-ball. Above a critical magnon number, the radial trapping potential for the magnons changes from harmonic to a "Mexican-hat" potential. The modification is eventually limited by the underlying profile of the magnetic field (see Fig.~\ref{Fig:qball-large}). Here, the systems' Hamiltonian mimics that of the $Q$-ball. All essential features of $Q$-balls, including the self-condensation of bosons into a spontaneously formed trap, long lifetime, and propagation in space across macroscopic distances (here several mm) have been demonstrated experimentally as shown in  Fig.~\ref{Fig:qball-large}. \cite{autti2018}

\subsection{Magnons-in-a-box -- The MIT Bag model}
\label{MIT}

The confinement mechanism of quarks in colorless combinations in quantum chromodynamics (QCD) is an open problem. One of the most successful phenomenological models, coined MIT bag model \cite{Hecht2000} as per the affiliation of its inventors, assumes a step change from zero potential within the confining region to a positive value elsewhere, a cavity surrounded by the QCD vacuum. The cavity is filled with false vacuum, in which the confinement is absent and quarks are free, thus creating the asymptotic freedom of QCD. Outside the cavity there is the true QCD vacuum, which is in the confinement phase, and thus a single quark cannot leave the cavity. Within the cavity quarks occupy single-particle orbitals, and there the zero point energy compensates the pressure from true vacuum.

\begin{figure}
\includegraphics[width=0.6\linewidth]{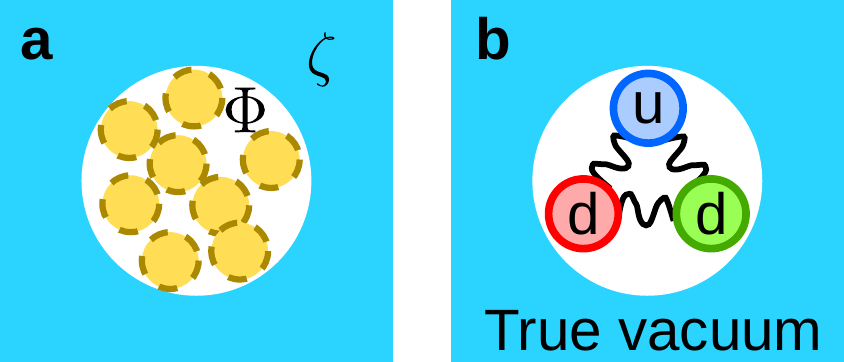}
\caption{{\bf Magnon BEC as a bosonic analogue to the MIT bag.} {\bf a} In the limit of large magnon density the magnon BEC carves a potential well (white), described by the charge field $\Phi (r)$, in the neutral field $\zeta$ which plays the role of the true vacuum. {\bf b} In the context of QCD, the quarks carve a potential well (false vacuum, white) in the true QCD vacuum, illustrated here for the charge-neutral neutron consisting of two down-quarks (labeled d) and one up-quark (labeled u).}
\label{Fig:MITBag}
\end{figure}

A similar situation is realized for a magnon BEC, if the magnetic maximum applied in the $Q$-ball experiment discussed in the previous Section is removed. Under these conditions the magnon BEC forms a self-trapping box analogously to the MIT bag model \cite{autti2012b,Autti2012}, c.f. Fig.~\ref{Fig:MITBag}. The flexible Cooper pair orbital momentum distribution $\hat{\bf l}$ plays either the role of the pion field or the role of the non-perturbative gluonic field, depending on the microscopic structure of the confinement phase. 

Much like quarks, magnons dig a hole in the confining "vacuum", pushing the orbital field away due to the repulsive interaction. The main difference from the MIT bag model is that magnons are bosons and may therefore macroscopically occupy the same energy level in the trap, forming a BEC, while in MIT bag model the number of fermions on the same energy level is limited by the Pauli exclusion principle. The bosonic bag becomes equivalent to the fermionic bag in the limit of large number of quark flavors due to bosonization of fermions. This phenomenon has been observed in cold gas experiments for ${\rm SU}(N)$ fermions \cite{PhysRevX.10.041053}.

\section{Light Higgs bosons: particle physics in magnon BEC}
\label{Higgs}

Both in the Standard Model (SM) of particle physics and in condensed matter physics the spontaneous symmetry breaking during a phase transition gives rise to a variety of collective modes. This is how the Higgs boson arises from the Higgs field in the Standard Model, for example. The gapless phase modes related to the breaking of continuous symmetries are called the Nambu-Goldstone (NG) modes, while the remaining gapped amplitude modes are called the Higgs modes. In superfluid $^3$He, we can make the magnon BEC interact with other collective modes, implementing scenarios that in the Standard Model context may require years of measurements using a major collider facility.

\begin{figure}
\includegraphics[width=1\linewidth]{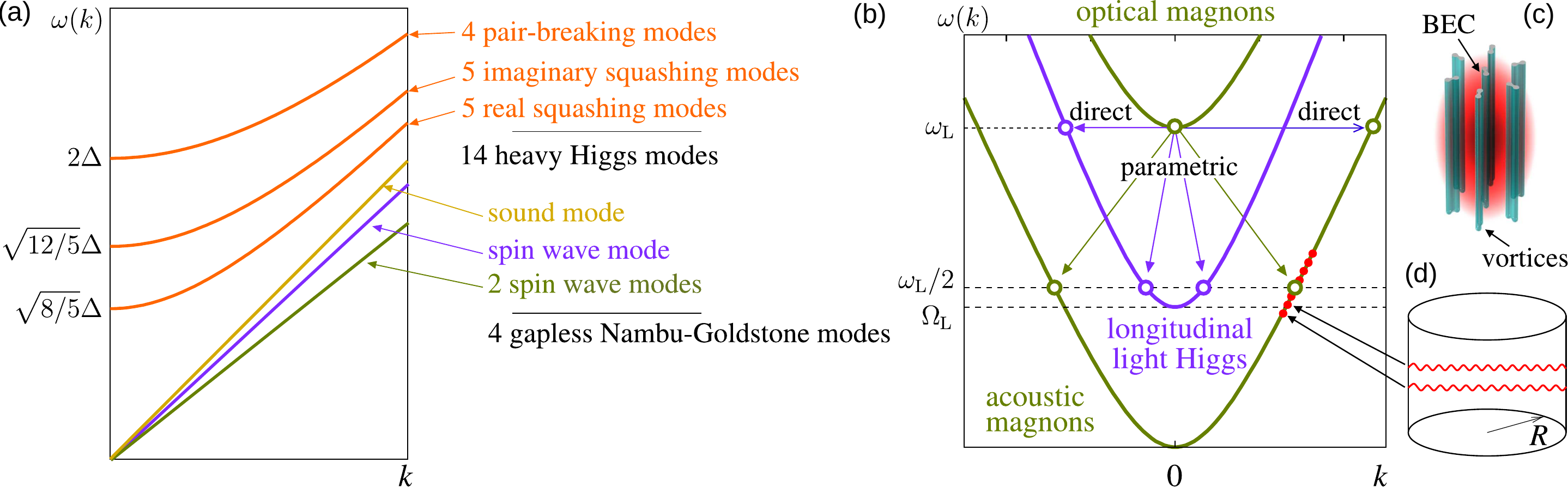}
\caption{{\bf Collective modes and decay channels.} (a) The collective mode spectrum in $^3$He-B contains six separate branches of collective modes. The 14 gapped Higgs modes (orange) are: four degenerate pair-breaking modes with gap $2 \Delta/h \sim 100\,$MHz, five imaginary squashing modes with gap $\sqrt{12/5}\Delta$, and five real squashing modes with gap $\sqrt{8/5} \Delta$. The gapless modes are a sound wave mode (oscillations of $\Phi$, yellow), a longitudinal spin wave mode (oscillations of $\theta$, purple), and two transverse spin wave modes (oscillations of $\hat{\bf n}$, green). (b) The longitudinal spin wave mode acquires a gap of $\Omega_{\rm L}/h \sim 100\,$kHz due to spin-orbit interaction and becomes a light Higgs mode. The transverse spin wave modes are split by the Zeeman effect in the presence of a magnetic field into optical and acoustic magnons. The arrows indicate possible decay channels. (c) The spatial extent of the optical magnon BEC in a typical experiment is of the order of a millimeter, and can be used as a probe for quantized vortices. (d) In a container of fixed size $R$, the spin wave modes form standing wave resonances.
\label{Fig:ModesAndDecayChannels}}
\end{figure}

The superfluid transition in $^3$He takes place via formation of Cooper pairs in the $L=1$, $S=1$ channel, for which the corresponding order parameter is a complex $3\times 3$ matrix combining spin and orbital degrees of freedom. Thus, $^3$He possesses 18 bosonic degrees of freedom, both massive (amplitude or Higgs modes) and massless (phase or Nambu-Goldstone modes), see Fig.~\ref{Fig:ModesAndDecayChannels}. The 14 Higgs modes have masses (energy gaps) of the order of the superfluid gap $\Delta/h \sim 100\,$MHz, where $h$ is the Planck constant, while the four Nambu-Goldstone modes (a sound wave mode and three spin wave modes) are massless at this energy scale. Higgs modes have been investigated for a long time theoretically \cite{ZeroSoundIn3HeB,10.1143/PTP.54.1,TEWORDT197697} and experimentally \cite{PhysRevLett.45.1952,PhysRevLett.61.1732,CollettJLTP2013} in $^3$He-B as well as $s$-wave superconductors \cite{PhysRevLett.111.057002,doi:10.1126/science.1254697} and ultracold Fermi gases \cite{BehrleHiggs}.

At low energies, the superfluid B phase of $^3$He breaks the {\it relative} orientational freedom of the spin and orbital spaces, and the resulting order parameter (at zero magnetic field) becomes
\begin{equation}
    A_{\alpha j} = \Delta e^{i \varphi} R_{\alpha j}(\hat{\bf n},\theta)\,,
\end{equation}
where the rotation matrix $R_{\alpha j}$ describes the relative orientation of the spin and orbital spaces; the spin space is obtained from the orbital space by rotating it by angle $\theta$ with respect to vector $\hat{\bf n}$. If the spin-orbit interaction is neglected or, equivalently, one considers energy scales of the order of the superfluid gap $\Delta$, the order parameter obtains an additional ("hidden") symmetry with respect to spin rotation. That is, the energy is degenerate with respect to $\hat{\bf n}$ and $\theta$.

The spin-orbit interaction lifts the degeneracy with respect to $\theta$, and the minimum energy corresponds to a rotation between the spin and orbital spaces by the Leggett angle $\theta_{\rm L} = \arccos (-1/4) \approx 104^{\circ}$. Due to this broken symmetry, one of the Nambu-Goldstone modes (the longitudinal spin wave mode) obtains a gap with magnitude equal to the Leggett frequency $\Omega_{\rm L}/h \sim 100\,$kHz. In the B-phase the longitudinal spin wave mode therefore becomes a light Higgs mode. Additionally, the presence of the magnetic field breaks the degeneracy of the transverse spin wave modes, one of which becomes gapped by the Larmor frequency $\omega_{\rm L} = |\gamma| H$, where $\gamma$ is the gyromagnetic ratio in $^3$He. The gapped transverse spin wave mode is called optical and the gapless mode acoustic. Throughout the manuscript, the term "magnon BEC" in the context of $^3$He-B refers to a BEC of optical magnons. 

Magnons in the BEC can be converted into other collective modes in the system. For example, the decay of the optical magnons of the BEC into light Higgs quasiparticles has been observed via a parametric decay channel in the absence of vortices \cite{HiggsNComm}, and via a direct channel in their presence\cite{PhysRevResearch.3.L032002} (see Sec.~\ref{sec:probe}) as illustrated in Fig.~\ref{Fig:ModesAndDecayChannels}. The parametric decay channel is directly analoguous to the production of Higgs modes in the Starndard Model. 

The separation of the Higgs modes in $^3$He-B into the heavy and light Higgs modes poses a question whether such a scenario would be realized in the context of the Standard Model as well. In particular, we note that the observed 125$\,$GeV Higgs mode\cite{Higgs1, Higgs2, Higgs3} is relatively light compared to the electroweak energy scale and, additionally, later measurements at higher energies \cite{DoubleHiggs2023} show another statistically significant resonance-like feature at the electroweak energies of $\approx 1\,$TeV related by the authors to possible Higgs pair-production. Entertaining the possibility of a $^3$He-B-like scenario, the observed feature could stem from formation of a "heavy" Higgs particle; in this case the 125~GeV Higgs boson would correspond to a pseudo-Goldstone (or a "light" Higgs) boson, whose small mass results from breaking of some hidden symmetry (see e.g. Ref.~\citenum{PhysRevD.92.055004} and references therein).

\section{Curved space-time: Event horizons}
\label{sec:spacetime}

\begin{figure}
\includegraphics[width=0.6\linewidth]{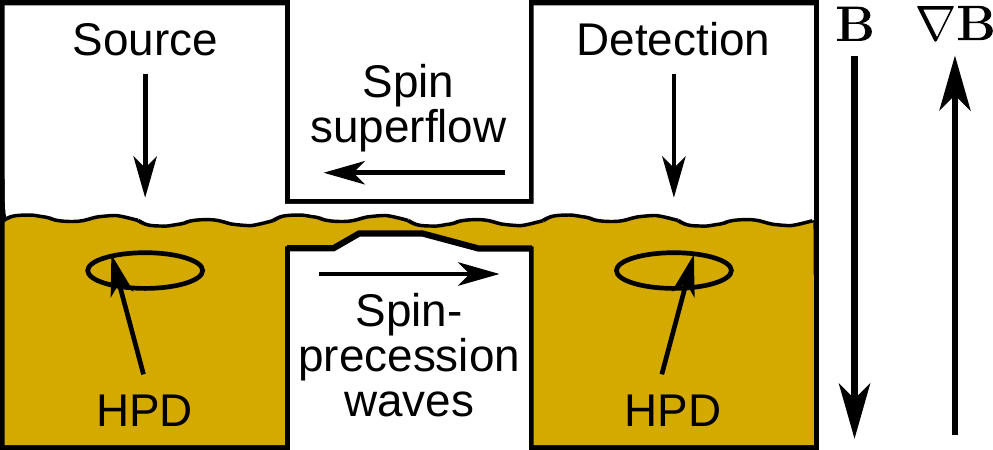}
\caption{{\bf Magnon BECs and event horizon.} In the experiment two volumes filled with superfluid $^3$He-B, in which the magnetization precesses uniformly (HPD) are connected by a narrow channel. An imposed phase difference between the precessing HPDs creates a spin supercurrent proportional to the phase difference. For sufficiently large magnitude of the spin superflow, counter-propagating spin-precession waves (surface-wave-like excitations of the HPD) can not propagate between the two volumes, analogously to the white-hole horizon. 
\label{Fig:Horizons}}
\end{figure}

The properties of the magnon BECs have also been utilized to study event horizons. In the conducted experiment \cite{PhysRevLett.123.161302}, two magnon BECs were confined by container walls and the magnetic field in two separate volumes connected by a narrow channel, Fig.~\ref{Fig:Horizons}. The channel contains a restriction, controlling the relative velocities of the spin supercurrents traveling in the bulk fluid and the spin-precession waves traveling along the surfaces of the magnon BEC.

The magnitude and direction of the spin superflow is controlled by the phase difference of the two magnon BECs, both of which are driven continuously by separate phase-locked voltage generators. The phase difference controls the spin supercurrents, while spin-precession waves are created by applied pulses. For a sufficiently large phase difference, spin-precession waves propagating opposite to the spin superflow are unable to propagate between the two volumes and instead are blocked by the spin superflow. This situation is analogous to a white-hole event horizon.

\section{Magnon BEC as a probe for quantized vortices}
\label{sec:probe}

Magnon BEC has proved itself as a useful probe of topological defects, especially quantized vortices. Vortices affect both the precession frequency of the condensate through modification of the trapping potential for magnons\cite{texture_vortex} and the relaxation rate of the condensate through providing additional relaxation channels \cite{Kondo199181,PhysRevResearch.3.L032002,PhysRevLett.127.115702,PhysRevResearch.2.033013}. Trapped magnon BECs provide a way to probe vortex dynamics locally and down to the lowest temperatures \cite{PhysRevB.97.014527, RQWT, Hosio_2013}, where there are still many open questions related to vortex dynamics, see e.g. Refs.~\citenum{doi:10.1146/annurev-conmatphys-031119-050821, Tsubota_2017}.

The effect of vortex configuration on the textural energy, which determines the radial magnon BEC trapping frequency, may be written as \cite{texture_vortex}
\begin{equation}
    F_{\rm v} = \frac{2}{5} a_{\rm m} H^{2} \frac{\lambda}{\Omega} \int {\rm d}^3 r \frac{\left( {\bm{\omega}}_{\rm v} \cdot \hat{\bf l} \right)^{2}}{\omega_{\rm v}}\,,
\end{equation}
where $a_{\rm m}$ is the magnetic anisotropy parameter, $\Omega$ is the angular velocity, $\lambda$ is a dimensionless parameter characterizing vortex contribution to textural energy, and $\bm{\omega} = \frac{1}{2}\langle \nabla \times {\bf v}_{\rm s} \rangle$ is the spatially averaged vorticity.

The vortex contribution to the textural energy may be introduced via a dimensionless parameter $\lambda$, which is contains the contributions from the orienting effect related to the superflow and the vortex core contribution. While the equilibrium vortex configuration is well understood \cite{doi:10.1126/sciadv.adh2899}, the mechanism of dissipation in the zero-temperature limit, in particular, remains an open problem.

\begin{figure}
\includegraphics[width=1\linewidth]{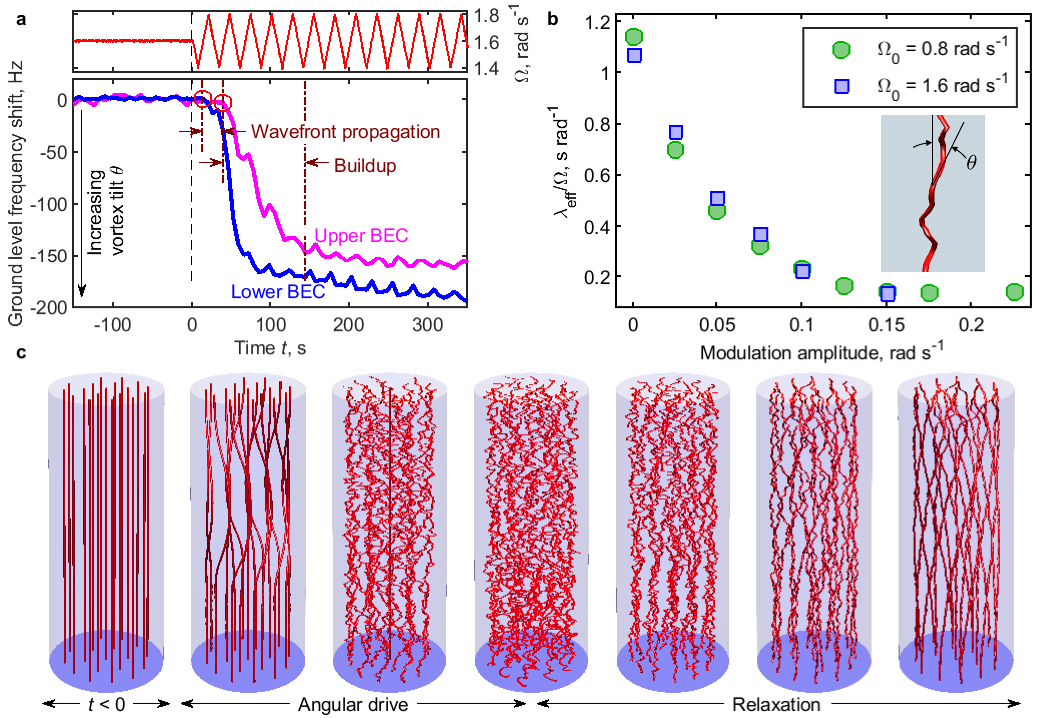}
\caption{{\bf Probing vortex dynamics with magnon BEC.} {\bf a} Vortex waves can be excited by applying perturbation in the form of angular modulation (top) on a steady vortex lattice. The vortex configuration is monitored locally with two separate magnon BECs ('Upper BEC' and 'Lower BEC'), which allows extracting time scales relevant for turbulence buildup. The angular drive results in decreased trapping frequency of the magnon BEC due to the $\omega_{\rm v \perp}$ term in Eq.\eqref{eq:lambdaeff}. {\bf b} As expected, the extracted value for $\lambda_{\rm eff}$ decreases monotonously with increasing drive amplitude. Both data are measured under the same experimental conditions with the same relative drive frequency $\omega/\Omega_0$, where $\Omega_0$ is the mean angular velocity during the drive. Inset shows the definition of the tilt angle $\theta$ with respect to the axis of rotation. {\bf c} Schematic illustration of the vortex (red) array evolution in response to the angular drive and the eventual relaxation after the drive is stopped.}
\label{Fig:LambdaSat}
\end{figure}

If the equilibrium vortex configuration is perturbed, i.e. $\bm{\omega} := {\bm \Omega} + {\bm \omega}'$, where ${\bm \Omega}$ is the equilibrium vorticity and $\bm{\omega}'$ is a random contribution with $\langle {\bm \omega}' \rangle = 0$, parameter $\lambda$ is replaced by an effective value
\begin{equation} \label{eq:lambdaeff}
    \lambda_{\rm eff} = \lambda \frac{1 + (\omega_{{\rm v} \parallel}/\Omega)^2 - (\omega_{{\rm v} \perp}/\Omega)^2}{\sqrt{1 + (\omega_{{\rm v} \parallel}/\Omega)^2 + 2(\omega_{{\rm v} \perp}/\Omega)^2}}\,.
\end{equation}
Here $\omega_{{\rm v} \parallel}$ and $\omega_{{\rm v} \perp}$ are the random contributions along the equilibrium orientation and perpendicular to it, respectively. This effect has been observed in experiments \cite{RQWT} by introducing vortex waves via modulation of the angular velocity around the equilibrium value and monitoring the precession frequency (i.e. the ground state energy) of the magnon BEC, see Fig.~\ref{Fig:LambdaSat}. Vortex core contribution can be extracted separately from the measured magnon energy levels by comparing measurements with and without vortices \cite{texture_vortex}.

\begin{figure}
\includegraphics[width=\linewidth]{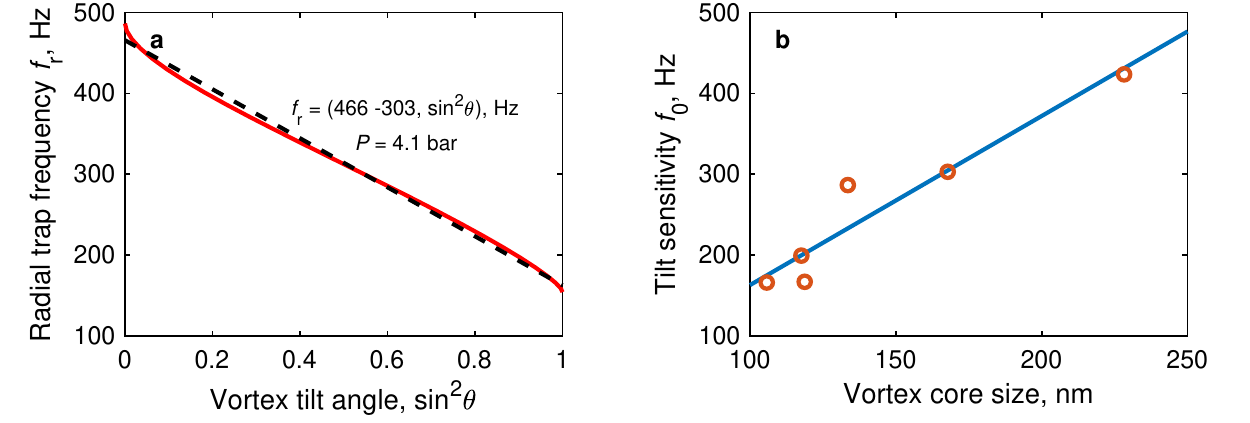}
\caption{{\bf Magnon BEC as a probe for vortex configuration.} {\bf a} The radial trapping potential for a magnon BEC, originating from the textural configuration in a cylindrical trap, scales with the vortex tilat angle roughly as $f_{\rm r} = \omega_{\rm r}/2 \pi \propto \sin ^2 \theta$, where $\theta$ is the tilt angle of vortices relative to the axis of rotation. The dashed line is a linear fit to the numerically calculated frequency shift using experimentally determined value for $(\lambda/\Omega)|_{\theta = 0}$ at 4.1 bar pressure (vortex core size $\sim 170\,$nm). The numerical model assumes uniform vortex tilt. {\bf b} The tilt sensitivity $f_0$, calculated using the measured $(\lambda/\Omega)|_{\theta = 0}$ at all pressures, is found to scale with roughly linearly with the vortex core size $(1+F_1^{\rm s}/3)\xi_0$, where $F_1^{\rm s}$ is the first symmetric Fermi-liquid parameter and $\xi_0$ is the $T=0$ coherence length.
}
\label{Fig:Sin2Prefactor}
\end{figure}

Based on numerical 1D calculations using the uniform vortex tilt model from Ref.~\citenum{kopu_texture}, where all vortices are tilted relative to the equilibrium position by the same angle, the magnon BEC ground state frequency, see Fig~\ref{Fig:Sin2Prefactor}{\bf a}, is found to scale as
\begin{equation} \label{eq:sin2prefactor}
    \Delta f \approx -f_0 \sin ^2 \theta\,,
\end{equation}
where the sensitivity $f_0 \sim 100\,$Hz is found to scale linearly with the vortex core size, see Fig.~\ref{Fig:Sin2Prefactor}{\bf b}. Using Eq.~\eqref{eq:sin2prefactor} one can then extract the average tilt angle of vortices within the volume occupied by the magnon BEC from the measured frequency shift. This method has been utilized for probing transient vortex dynamics \cite{RQWT}.

When quantized vortices penetrate the magnon BEC, like in Fig.~\ref{Fig:ModesAndDecayChannels}c, they also contribute to enhanced relaxation of the condensate. Distortion of the superfluid order parameter around vortex cores opens direct non-momentum-conserving conversion channels of optical magnons from the condesate to other spin-wave modes, predominantly light Higgs\cite{Laine2018}, see Fig.~\ref{Fig:ModesAndDecayChannels}b. The decay rate of macroscopic (mm-sized) condensate depends on the internal structure of microscopic (100\,nm-sized) vortex core. This effect has been used in experiments to distinguish between axially symmetric and asymmetric double-core vortex structures\cite{Kondo199181} and to measure the vortex core size \cite{PhysRevResearch.3.L032002}.

\section{Outlook}
\label{sec:outlook}

Bose-Einstein condensates of magnon quasiparticles have been realized experimentally in different systems including solid-state magnetic materials, dilute quantum gases, and superfluid $^3$He. Spin superfluidity of those condensates and phenomena such as the Josephson effect may be viewed as analogs of superconductivity in the magnetic domain. Superconducting quantum electronics, based on Josephson junctions, is one of the most important platforms for quantum technologies, however requiring dilution refrigerator temperatures of about 10\,mK. Coherent magnetic phenomena are generally more robust to temperature than superconductivity with existing room-temperature demonstrations of magnon BEC, spin supercurrents and the magnon Josephson effect\cite{Demokritov2006,Bozhko20161057,PhysRevB.104.144414}. Thus one of the strong axes of research on magnon condensates is the development of practically useful (super-)magnonic devices operating at ambient conditions \cite{Serga_2010, 9706176, AiM, doi:10.1021/acsami.1c01562}. In this Perspective, we have shown that there is another important dimension of magnon BEC applications as a laboratory to study fundamental questions in various areas of physics from Q-balls and Higgs particles to time crystals and quantum turbulence. These fundamental phenomena can and should be utilized in future magnon-based devices.

Advances outlined in this Perspective have triggered suggestions for further applications of magnon BEC for fundamental research. In particular, magnon BEC has been proposed to be used as a dark matter axion detector \cite{PhysRevLett.129.211801, foster2023statistics} via the coupling between axions and coherently precessing spins. The coupling gives rise to a term in the Hamiltonian of the BEC that looks like an effective magnetic field, oscillating with the frequency depending on the axion mass. When the frequency of precession matches the frequency of the axion field, potentially observable glitches in the magnon BEC decay are expected. The change of the frequency of the magnon BEC during decay due to inter-magnon interactions (see an example in Fig.~\ref{Fig:qball-large}a) allows to probe a continuous range of the axion masses.

Magnon BECs could also be used as a source and a detector of spin currents in particular to probe composite topological matter at the interface of superfluid $^3$He and graphene \cite{Katsnelson_2014}. Atoms of $^3$He do not penetrate through a graphene sheet, but coupling of the spin degrees of freedom of two superfluids on the opposite sides of the sheet immersed in the liquid is nevertheless possible through the excitations of the graphene itself (electronic or ripplons) or via magnetic coupling of the quasiparticles living at the interface between graphene and helium superfluid, including Majorana surface states. As in the original observation of the spin Josephson effect in $^3$He \cite{BR2}, two magnon condensates separated by a channel can be maintained at the controlled phase difference of the magnetization precession, which drives a spin current through the channel similar to Fig.~\ref{Fig:Horizons}. In this case, instead of the constriction, one would place a graphene membrane across the channel and find whether the Josephson coupling is nevertheless observable.

Even a single magnon BEC placed in contact with the surface of topological superfluid can provide valuable information on the topological surface states. Trapping a condensate with the magnetic field profile, like in Fig.~\ref{Fig:ExpSetup}, allows to move the BEC around the fluid simply by adjusting the trapping field. Preliminary measurements show that magnon loss from BEC is significantly enhanced when the condensate is brought from bulk to the surface of $^3$He-B sample \cite{PhysRevB.86.024506}. Future experiments should clarify whether this relaxation increase is caused by Majorana surface states \cite{PhysRevLett.103.235301,Silaev_2018}. The ability to manipulate Majorana states would come with applications in quantum information processing.

Development of optical lattices opened many new areas of physics for probing in cold atom experiments \cite{BlochReview}. So far experiments on magnon BECs were limited to one or two condensates. An exciting development will be to form magnon condensates on a lattice to probe solid-state physics in magnetic domain, perhaps utilizing spinor cold gases \cite{Stamper-Kurn20131191} on a 2D lattice of elongated trapping tubes \cite{PhysRevLett.116.095301} created by superimposing two orthogonal standing waves of attractive lase beams, or in solid-state systems where one may similarly utilize optical beamshaping techniques to directly print a 2D lattice and impose spin currents \cite{YIGSpinCurrMagnonBEC}. For magnon BEC in superfluid $^3$He, multiple regularly arranged traps can potentially be formed using orbital part of the trapping potential, Fig.~\ref{Fig:ExpSetup}. Array of the quantized vortices, created by rotation, modulate the orbital degrees of freedom of superfluid $^3$He and at certain conditions may form individual magnon traps around the vortex cores \cite{Doublequantum}. The spacing of the lattice sites can then be regulated by rotation velocity. This will change coupling between condensates at individual sites which will allow to observe, for example, a superfluid-Mott insulator transition\cite{SuperfluidMott} but for spin superfluid.

Excitations of the magnon BEC, in particular its Nambu-Goldstone mode (phonon of a time crystal) provide ample possibilities to model propagation of particles in curved space in acoustic-metric type experiments \cite{Visser19981767}. In such models, effective metric is created by the fluid flow which is externally controlled, while the dispersion relation of the propagating modes is usually nature-given, like gravity waves on water \cite{Schutzhold2002, Weinfurtner2011, PhysRevLett.126.041105}. For magnon BEC, remarkably, the spectrum of Goldstone bosons can typically be adjusted in a wide range by external magnetic field, while the spin flow is formed by the phase of the coherent precession controlled with rf pumping. Thus non-trivial metrics can be realized even without using geometrical flow conditioners (like a channel in Fig.~\ref{Fig:Horizons}) and cases impossible for phonons or ripplons in classical fluids can be achieved. For example, in a magnon BEC in the polar phase of $^3$He, the propagation speed of the Goldstone boson is controlled by the angle of the static magnetic field with the orbital anisotropy axis of the superfluid. The mode can be brought to a complete halt at a critical angle, and beyond this angle the metric changes signature from Minkowski to Euclidean \cite{Nissinen_2017}. Using magnon BEC one can potentially study instability of quantum vacuum in such signature transition.

Magnon BECs make one of the most versatile implementations of time crystals that also comes the closest to the ideal time crystal of all systems in the laboratory. Expanding on the experiments summarized in this Perspective, one may pose fundamental questions such as is it possible to melt a time crystal into a time fluid, is it possible to seed time crystallization \cite{Hajdusek2022}, or how time crystals interact with different types of matter. The time crystal description of magnon BECs also emphasizes potential for quantum magnonics applications: the magnitude and phase of the wave function of a single magnon-BEC time crystal, or that of a multi-level composite system of time crystals, is directly accessible in experiments, revealing basic quantum mechanical processes such as Landau-Zener transitions and Rabi oscillations in a non-destructive measurement in real time. These can therefore be harnessed unimpeded for also technological applications.

Additionally, physics similar to (and beyond!) that outlined in this perspective can perhaps be studied in systems for which the experimental realization is yet to come. One promising system is the superfluid fermionic spin-triplet quantum gas, which could be realized by synthetic gauge fields e.g. through Rashba-coupling scenarios \cite{Manchon_2015}, by tuning into a $p$-wave Feshbach resonance \cite{Atompwave}, or perhaps by induced interactions \cite{PhysRevLett.121.253402}. In the context of the weak-coupling theory, the B-like phase is always expected to be the lowest energy superfluid phase \cite{vollhardt2013superfluid}. Therefore, it is expected that the spin-orbit interaction opens up a gap for one of the Goldstone modes, giving rise to the light Higgs mode. We note that such a scenario allows for unique research directions as the spin-orbit coupling strength is likely controllable e.g. via the Rashba coupling strength, via the amplitude of magnetic field, or via density of the inducing component, depending on the experimental setup. 

Another interesting research project would be to study the properties of magnon BECs in a (putative) spin-triplet superconductor, such as UTe$_2$, see e.g. Ref.~\citenum{Aoki_2022} and references therein. The order parameter of UTe$_2$ may take multiple forms, including the one whose $d$-vector representation is $\hat{\bf d}({\bf k}) \propto (0,k_y,k_z)$. Such an order parameter corresponds to the $B_{3g}$ irreducible representation of the $D_{2h}$ point symmetry group in UTe$_2$ and to the so-called planar phase\cite{vollhardt2013superfluid} in the context of $^3$He. In $^3$He the planar phase is predicted to never be the lowest energy phase, as its energy always lies between the B phase and the polar phase \cite{vollhardt2013superfluid}. Due to the presence of the discrete point symmetry group, similar argumentation may not apply in UTe$_2$, making this a unique possibility to study yet another novel topological phase, including its collective modes such as spin waves and by extension the magnon BEC. The strength of the spin-orbit coupling in UTe$_2$ remains an open question \cite{Aoki_2022}, but it is expected to be non-zero and quite possibly significant (bare uranium has a large spin-orbit interaction strength). As long as the spin-orbit coupling is non-zero, a gap opens in the longitudinal magnon spectrum which then becomes a light Higgs mode. In principle, the respective order parameter is also unique in that it supports isolated monopoles, i.e. monopoles that do not act as termination points for linear objects such as Dirac monopoles \cite{VOLOVIK2022168998}.

To conclude, magnon BECs are interesting systems in their own right, as they form analogies to various fields of physics, from time crystals and particle physics to QCD, and provide non-invasive methods for probing the dynamics and structure of quantized vortices. Moreover, magnon BECs hold enormous future potential for accessing novel physics, as replacements for electronic components, or perhaps for detecting dark matter.

\section*{Data availability}
Data sharing is not applicable to this article as no new data were created or analyzed in this study.
\section*{Author Declarations}
The authors have no conflicts to disclose.

\begin{acknowledgments}
We thank Grigory Volovik for stimulating discussions. The work was supported by Academy of Finland grant 332964 and by UKRI EPSRC grant EP/W015730/1. 
\end{acknowledgments}

\bibliography{magnon_bibliography}

\end{document}